\documentclass[final,3p,twocolumn]{elsarticle}

\usepackage{graphicx}

\usepackage{amssymb}

\journal{Physica C}

\begin{document}

\begin{frontmatter}



\title{Consequences of the peculiar intrinsic properties of MgB$_2$ on its macroscopic current flow}

\author{M. Eisterer}
\ead{eisterer@ati.ac.at}
\author{H. W. Weber}
\address{Atominstitut, Vienna University of Technology, 1020 Vienna, Austria}

\begin{abstract}
The influence of two important features of magnesium diboride on
the macroscopic transport properties of polycrystalline MgB$_2$ is
discussed in the framework of a percolation model. While two band
superconductivity does not have significant consequences in the
field and temperature range of possible power applications, the
opposite is true for the anisotropy of the upper critical field.
The field dependence of the critical current densities strongly
increases and the macroscopic supercurrents disappear well below
the apparent upper critical field. The common scaling laws for the
field dependence of the volume pinning force are altered and
Kramer's plot is no longer linear, although grain boundary pinning
dominates in nearly all polycrystalline MgB$_2$ conductors. In
contrast to the conventional superconductors NbTi and Nb$_3$Sn, a
significant critical current anisotropy can be induced by the
preparation technique of MgB$_2$ tapes.

\end{abstract}


\end{frontmatter}

This contribution focuses on the peculiar properties of magnesium
diboride, MgB$_2$, and their influence on the macroscopic current
transport: two band superconductivity, the anisotropy of the upper
critical field, $\gamma$, and the strong effect of impurity
scattering. The two types of charge carriers in the $\sigma$- and
$\pi$-bands of the boron orbitals \cite{An01} have different
energy gaps \cite{Maz03} and their contributions to the superfluid
density changes differently when a magnetic field is applied
\cite{Eis07rev}. While both bands contribute approximately equally
at zero field, the superconducting charge carriers of the
$\pi$-band are rapidly suppressed by small magnetic fields. This
leads to a field and temperature dependence of the magnetic
penetration depth \cite{Gol02,Ang04,Eis05b} and significant
alterations of the reversible magnetization \cite{Kle06,Eis05b}.
However, even at zero field, the contribution of the $\pi$-band to
the condensation energy \cite{Eis05b} and to the critical current
is small ($\sim 20$\%) \cite{Nic05b,Eis07rev}, which is a
consequence of the small energy gap in the $\pi$-band. This
contribution is further suppressed by the application of rather
small magnetic fields. This means that the theoretically
interesting two band nature of superconductivity in MgB$_2$ has a
negligible influence on the current transport under application
relevant operation conditions and the material can be modeled as a
single ($\sigma$-)band superconductor.

The influence of the intrinsic anisotropy on the macroscopic
transport properties is much more significant. Polycrystalline
materials with randomly oriented grains can be considered as
inhomogeneous in magnetic fields \cite{Eis05}. Different grains
attain different properties, defined by their orientation with
respect to the applied field. This results in a strong field
dependence of the critical current density, $J_\mathrm{c}$, even
in the case of strong pinning \cite{Eis03}. At the zero
resistivity (or irreversibility) field, $B_{\rho=0}$, above which
currents are no longer loss free, the remaining superconducting
volume of the sample decomposes into disconnected clusters of
superconducting grains. $B_{\rho=0}$ is significantly below the
apparent upper critical field, $B_\mathrm{c2}$, which is defined
by grains oriented with their boron planes parallel to the applied
field and corresponds to $B_\mathrm{c2}^{ab}$. It is approximately
14\,T in the clean limit \cite{Eis07rev}. However, large currents,
as required for most applications, can be expected \cite{Eis05}
only up to about $B_\mathrm{c2}^{c}=B_\mathrm{c2}^{ab}/\gamma$,
which is $\sim$3\,T in clean MgB$_2$. Fortunately, $B_\mathrm{c2}$
can be significantly enhanced by the introduction of disorder
\cite{Put05,Eis07,Gur04}, which also reduces the upper critical
field anisotropy \cite{Kru07}. Both effects enlarge the maximum
operation field. This improvement occurs at the expense of a
$T_\mathrm{c}$ reduction \cite{Put07}, which competes with the
beneficial effects of disorder and even leads to a degradation of
the properties at high temperatures. Hence, material optimization
for operation at$~\sim 20$\,K is much more difficult than for
operation in liquid helium.

\begin{figure} \centering \includegraphics[clip,width=0.4\textwidth]{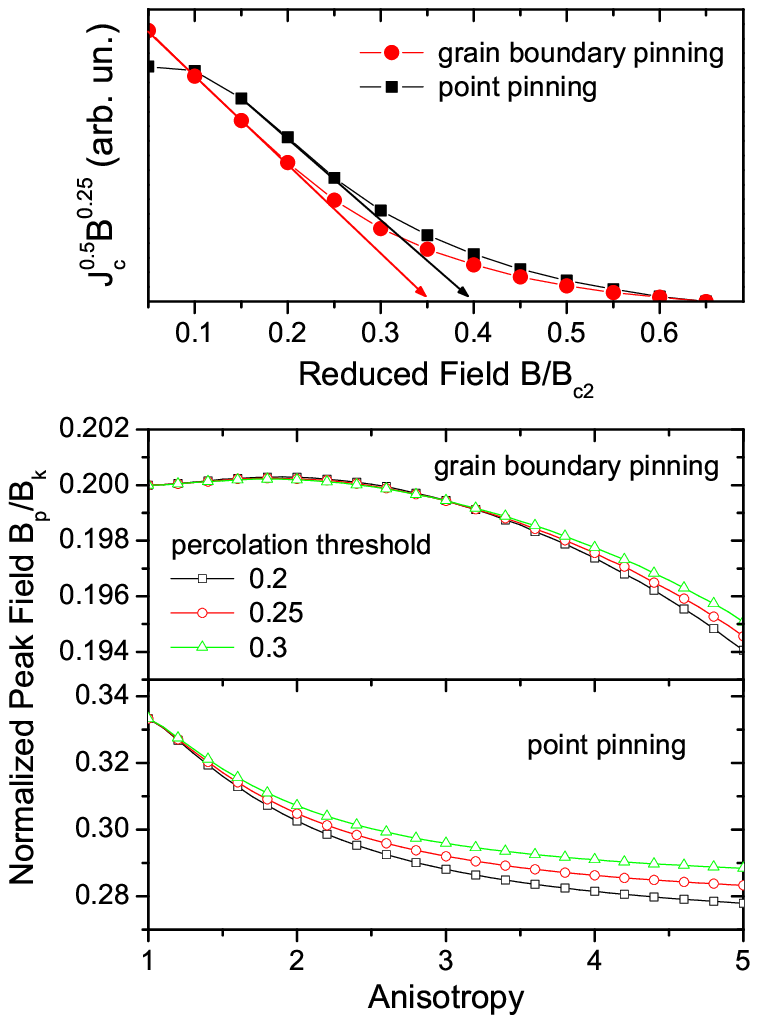}
\caption{Determination of the Kramer field $B_\mathrm{k}$.}
\label{Fig1}
\end{figure}

\begin{figure} \centering \includegraphics[clip,width=0.4\textwidth]{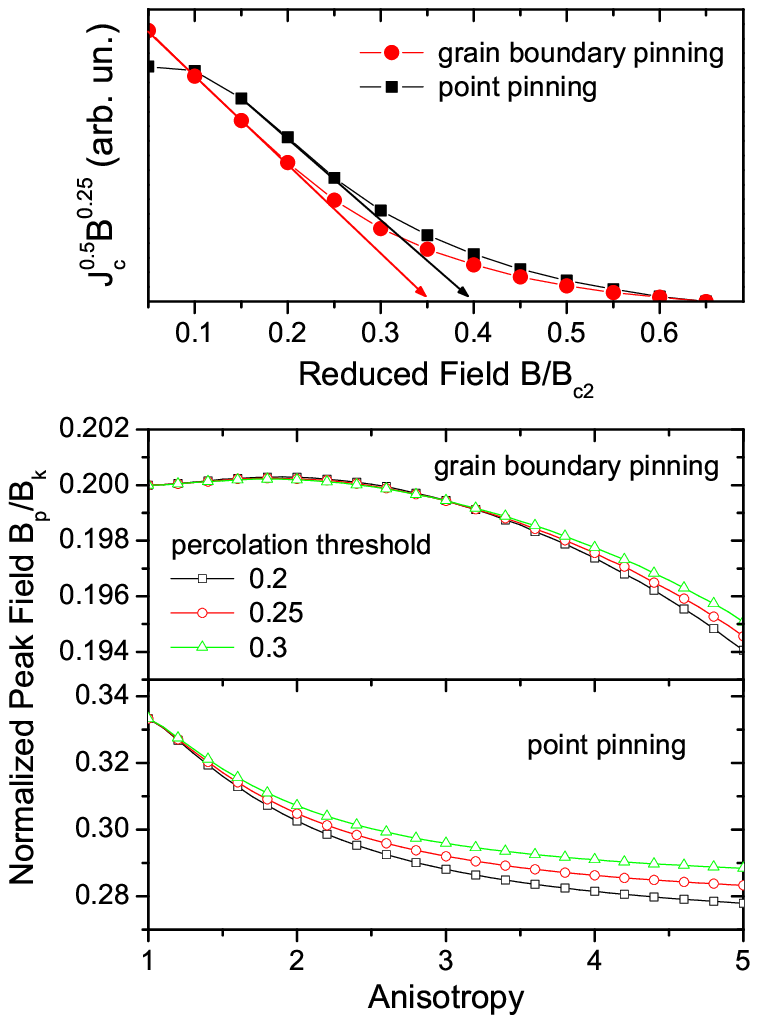}
\caption{The peak position $B_\mathrm{p}$ does not significantly
change with anisotropy, if the field is normalized by the Kramer
field.} \label{Fig2}
\end{figure}

The changed field dependence of the critical currents also alters
the Kramer plot and shifts the peak of the volume pinning force,
$F_\mathrm{p}(B)$, to significantly lower fields \cite{Eis08}. The
Kramer plot is no longer linear near the irreversibility field and
extrapolation of its (pseudo-)linear part near the inflection
point (Fig.~\ref{Fig1}) results in a Kramer field $B_\mathrm{k}$,
which is in between $B_\mathrm{c2}^{c}$ and $B_{\rho=0}$. It is
somewhat surprising that the peak in $F_\mathrm{p}(b)$ remains
nearly at the position for isotropic superconductors, if $B$ is
normalized by $B_\mathrm{k}$ instead of $B_{\rho=0}$ \cite{Mar07}.
(They are equivalent only in isotropic superconductors and the
Kramer plot was originally proposed for the determination of
$B_{\rho=0}$.) The small influence of $\gamma$ is demonstrated for
grain boundary (upper panel) and point pinning (lower panel) in
Fig.~\ref{Fig2}. Even $F_\mathrm{p}(b)$ follows the theoretical
isotropic behavior \cite{Dew74} up to $b=B/B_\mathrm{k}=1/2$.

The macroscopic critical current anisotropy observed in some
MgB$_2$ tapes \cite{Lez06b,Kov08} is another manifestation of the
intrinsic anisotropy. The rolling process induces partial texture,
which renders the distribution function of the local
$J_\mathrm{c}$ dependent on the orientation of the tape with
respect to the field \cite{Eis09p}. This effect is absent in
conventional isotropic superconductors as Nb$_3$Sn and NbTi. The
cuprates on the other hand, need a much higher degree of texture,
since large angle grain boundaries strongly reduce the macroscopic
current. Hence, all grains must be oriented nearly identically and
$J_\mathrm{c}$ should be, therefore, equal in all of them.





\bibliographystyle{elsarticle-num}





\end{document}